\documentclass[prd,a4paper,twocolumn,preprintnumbers,nofootinbib,superscriptaddress]{revtex4}

\pdfoutput=1
\usepackage[english]{babel}
\usepackage{amsmath,amssymb,amsfonts, bm,bbm,slashed, subdepth}
\usepackage{graphicx}
\usepackage{enumerate}
\usepackage{setspace}
\usepackage{booktabs, tabularx}
\usepackage{units}
\usepackage{color}
\usepackage{multirow}
\usepackage[dvipsnames]{xcolor}
\usepackage[pscoord]{eso-pic}
\usepackage[normalem]{ulem}

\usepackage{scalerel}

\usepackage{hyperref}
\hypersetup{ 
 setpagesize=false,
 bookmarksnumbered=true,
 colorlinks=true,
 linkcolor=blue,
 citecolor=red,
 hypertexnames=true
}

\newcommand{\be}{\begin{equation}}
\newcommand{\ee}{\end{equation}}
\newcommand{\ba}{\begin{array}}
\newcommand{\ea}{\end{array}}
\newcommand{\bea}{\begin{eqnarray}}
\newcommand{\eea}{\end{eqnarray}}
\newcommand{\balg}{\begin{align}}
\newcommand{\ealg}{\end{align}}
\newcommand{\bit}{\begin{itemize}}
\newcommand{\eit}{\end{itemize}}

\makeindex

\begin{document}
\preprint{ULB-TH/20-19}

% =============================================================================
\title{
Neutrino lines from DM decay induced by high-scale seesaw interactions
}
% =============================================================================

\author{Rupert Coy}
\email{rupert.coy@ulb.ac.be}
\affiliation{Service de Physique Th\'eorique, Universit\'e Libre de Bruxelles, Boulevard du Triomphe, CP225, 1050 Brussels, Belgium}

\author{Thomas Hambye}
\email{thambye@ulb.ac.be}
\affiliation{Service de Physique Th\'eorique, Universit\'e Libre de Bruxelles, Boulevard du Triomphe, CP225, 1050 Brussels, Belgium}

\begin{abstract}
If the stability of the dark matter (DM) particle is due to an accidental symmetry, nothing prevents UV physics from destabilising it by inducing DM decays suppressed by powers of the UV scale. The seesaw physics, presumably at the origin of neutrino mass,
could induce such a decay. We show that if the seesaw scale lies around the usual Weinberg operator scale, the induced DM decay could generically lead to neutrino lines whose intensity is of the order of the present sensitivity of neutrino telescopes. 
We illustrate this possibility with models in which the DM is made of the gauge boson(s) of an abelian or non-abelian gauge symmetry. 
\end{abstract}

\maketitle
%==============================================================================

%%%%%%%%%%%%%%%%%%%
\section{Introduction}

The four stable particles that exist in the Standard Model (SM) are all stable for a fundamental reason, related to Lorentz invariance, the gauge symmetries of the SM and the quantum numbers of the SM particles under these gauge symmetries. The evidence for a fifth (or more) stable particle, the dark matter (DM) particle(s), raises the question of whether its stability hides a new fundamental symmetry/principle, for instance a new gauge symmetry, rather than just an (often assumed) ad hoc discrete symmetry (see e.g.~\cite{Hambye:2010zb}). Among various stabilisation mechanisms, the possibility that the DM particle(s) would be stable due to an accidental symmetry is rather intriguing. Various frameworks of this type can be considered. One option is simply to assume that DM belongs to a large enough weak multiplet that no renormalisable interactions that could destabilise it can be written down \cite{Cirelli:2005uq}. Another option consists of assuming a new gauge symmetry whose breaking leaves an accidental symmetry which is not a subgroup of the gauge symmetry. This can be done on the basis of an abelian \cite{Hambye:2008bq,Lebedev:2011iq} or non-abelian gauge symmetry \cite{Hambye:2008bq,Hambye:2009fg,Arina:2009uq,Gross:2015cwa,Hambye:2018qjv,Buttazzo:2019iwr,Buttazzo:2019mvl}. Other possibilities of course do exist.

If DM is accidentally stable, nothing forbids some UV physics, lying at the scale $\Lambda_{UV}$, from destabilising it. 
This is similar to what is expected for the proton, for instance, in GUT theories. 
The DM lifetime must obviously be longer than the age of the universe, $\sim 10^{18}$~sec, or in fact much longer, $\tau_{DM}\gtrsim 10^{22-29}$~sec, in order not to produce fluxes of cosmic rays larger than those observed (assuming decays into SM particles, depending on the decay channel and DM mass, and considering here $m_{DM}\gtrsim 1$~GeV). 
If the UV physics induces a decay amplitude which is suppressed by only one power of the UV scale, the DM decay width induced is typically proportional to $(1/8\pi)\cdot m^3_{DM}/\Lambda^2$, and it is known that this is many orders of magnitude too fast to fulfil these constraints (for $\Lambda$ no larger than the Planck scale). Tiny couplings are necessary in this case, so that $\Lambda$ is an effective scale quite different from the much lower fundamental scale $\Lambda_{UV}$, i.e.~$\Lambda=\Lambda_{UV}/g$, where $g \ll 1$ is some combination of couplings. Instead, a decay amplitude suppressed by two powers of the UV scale, $\Lambda=\Lambda_{UV}$, gives a decay width proportional to $(1/8\pi)\cdot m^5_{DM}/\Lambda_{UV}^4$. 
For DM mass of order the electroweak scale and $\Lambda_{UV}$ of order the GUT scale,
this nicely leads to lifetimes of order the lower bound from cosmic rays.\footnote{For instance, for $m_{DM}=100$~GeV and $\Lambda=10^{15}$~GeV, one gets $\tau_{DM}\sim 10^{27}$~sec. More generally, this holds replacing $m^5_{DM}$ in the decay width with any dimension-5 combination of masses around the electroweak scale.} In this case, there is the possibility of a direct connection between the fundamental UV scale and the DM lifetime.
Besides neutrino mass and proton decay probes, this provides another nice avenue to study very high scale physics, which we investigate in this letter.

Probably the most motivated UV physics one could consider to destabilise the DM particle is the seesaw physics. 
Although the seesaw states, for instance right-handed neutrinos $N_i$ in the type-I seesaw model, could lie at a low scale, clearly the smallness of the neutrino masses fits very well with these seesaw states being at a much higher scale than the electroweak scale, not far from the GUT scale. For seesaw Yukawa couplings of order unity, the seesaw scale is the scale of the $LLHH/\Lambda_W$ Weinberg operator, $\Lambda_W\sim 10^{15}$~GeV.
The seesaw interactions are not necessarily expected to cause relevant DM decays. 
For instance, adding right-handed neutrinos to the minimal DM quintuplet $\psi_{DM}^{(5)}$ setup \cite{Cirelli:2005uq} doesn't easily induce a decay of this quintuplet.\footnote{The lowest-dimensional operator involving both fermions is $\bar{N}H^4\psi_{DM}^{(5)}/\Lambda^3$, which from the exchange of a right-handed neutrino leads to the dimension-8 operator, $\bar{L}H^5\psi_{DM}^{(5)}/(\Lambda^3 m_N)$.}
%Thus, the seesaw interactions do not destabilise the quintuplet. 
However, there are other models where the seesaw is expected to cause such a decay, see \cite{Rothstein:1992rh,Berezinsky:1993fm,Lattanzi:2007ux,Bazzocchi:2008fh,Frigerio:2011in,Lattanzi:2013uza,Queiroz:2014yna,Dudas:2014bca,Wang:2016vfj,Garcia-Cely:2017oco,Patel:2019zky} and below.
In particular, if the DM, and more generally its associated sector, is comprised of SM singlet particles, the right-handed neutrinos can easily couple to this sector, also being SM singlets. This allows the decay $DM\rightarrow N^*+X$, with a further conversion of the virtual right-handed neutrino, $N^*$, into a SM neutrino through seesaw mixing,
thereby allowing DM to decay into neutrinos.
The seesaw interactions therefore not only offer the possibility of inducing a slow DM decay, but also a way of easily producing SM neutrino(s) in the final state, in particular a neutrino line if the decay is to a two-body final state. As is well known, monochromatic $\gamma$ \cite{Srednicki:1985sf,Bergstrom:1988fp,Rudaz:1989ij,Bouquet:1989sr,Rudaz:1990rt,Bergstrom:1997fj,Gustafsson:2013gca} or neutrino \cite{Baratella:2013fya,Dudas:2014bca,Aisati:2015vma,Aisati:2015ova,ElAisati:2017ppn} signals are ``DM smoking guns" because there is basically no astrophysical background for such a signal.

From the discussion above, it is clear that if, through the exchange of a heavy seesaw state, an operator (that is, a decay amplitude) suppressed by only one power of the seesaw scale is generated, the decay 
will naturally be far too fast, unless the DM mass scale is quite low (well below the GeV scale) and/or this seesaw exchange diagram involves small couplings or extra tiny mass ratios. 
%{\bf Along} This isn't the right word
In all these ways out, the direct connection between the Weinberg operator scale and the DM lifetime is lost. 
This situation occurs for instance in Majoron DM models \cite{Chikashige:1980ui,Schechter:1981cv,Rothstein:1992rh,Berezinsky:1993fm,Lattanzi:2007ux,Bazzocchi:2008fh,Frigerio:2011in,Lattanzi:2013uza,Queiroz:2014yna,Wang:2016vfj,Garcia-Cely:2017oco}, in which the decay amplitude into a pair of charged leptons, suppressed by only one power of the seesaw scale, is induced at the one-loop level.
Another example of this situation was recently considered in \cite{Patel:2019zky}.
However, if a model manages to not induce any decay amplitude suppressed by one power of the seesaw scale, but does induce an amplitude suppressed by two powers of this scale, the direct connection between neutrino mass and the DM lifetime can hold. 
This moreover leads to a neutrino line with intensity of the order of the sensitivity of present indirect detection experiments. This is the possibility we consider in this work.
%

%In the following we show that
%other simple frameworks nevertheless do lead to the desired pattern of a decay amplitude dominated by a term suppressed by two powers of the seesaw scale.

%There are 2 reasons why seesaw interactions can easily destabilize the DM particle(s). On the one hand this is due to the fact that right-handed neutrinos can couple to the DM sector because they are SM singlets. On the other hand this is due to the fact that the virtual right-handed neutrino(s) coupling to the DM particle(s) can convert themselves into SM neutrinos through $N-\nu_L$ seesaw mixing. 

%%%%%%%%%%%%%%%%%%%%%%%%
\section{A simple setup}

The example model we will consider in detail assumes an extra $U(1)_X$ gauge symmetry spontaneously broken by the vacuum expectation value of a scalar boson, $\phi$, with the addition of a vector-like fermion charged under it, $\chi$.\footnote{A more involved chiral fermion structure in which the fermions acquire their mass from the spontaneous breaking of a gauge symmetry could also be considered.} 
The associated Lagrangian is
\begin{eqnarray}
{\cal L}&=&{\cal L}_{SM}-\frac{1}{4} F^X_{\mu\nu}F^{X\mu\nu}+\bar{\chi} (i D\hspace{-2.1mm}\slash \hspace{1.1mm}-m_\chi) \chi+D_\mu \phi^\dagger D^\mu \phi\nonumber\\
&&-\lambda_m\phi^\dagger\phi H^\dagger H -V(\phi) \, ,
\label{hiddenvectorLagr}
\end{eqnarray}
where $D_\mu=\partial_\mu-i g_X Q_X A'_\mu$, $V(\phi)=\mu^2 \phi^\dagger \phi +\lambda_\phi (\phi^\dagger \phi)^2$, and $F^X_{\mu \nu}$ is the $U(1)_X$ field strength tensor. Here we assume that there is no kinetic mixing interaction between the $U(1)_X$ and hypercharge gauge bosons. 
%It cannot subsequently be generated at loop-level (so indeed a tiny kinetic mixing parameter is technically natural). 
We parameterise the scalar by $\phi = (\eta' + v_\phi)/\sqrt{2}$, with $v_\phi = \sqrt{-2\mu^2/\lambda_\phi}$, the NGB from the spontaneously broken $U(1)_X$ being eaten by the $A'$.  Without the fermion $\chi$, this is the DM model of Refs.~\cite{Hambye:2008bq,Lebedev:2011iq}, where the $U(1)_X$ gauge boson, $A'$,
is the DM candidate. It is stable because after spontaneous breaking, the model displays an accidental $\mathbb{Z}_2$ symmetry under which the gauge boson is odd.\footnote{Actually, unlike for the non-abelian case, this charge conjugation symmetry of the abelian case is not fully accidental here since it holds only if one assumes no kinetic mixing.}
%As is known, for a non-abelian gauge symmetry instead, the remnant symmetry is fully accidental, see below.
Adding the extra fermion, $\chi$, leads to two possible DM patterns. If $m_{A'} =g_X v_\phi >2m_\chi$, the vector boson decays into a pair of fermions and is not stable anymore, thanks to the fact that the fermion-gauge boson interaction breaks the $\mathbb{Z}_2$ symmetry. But the $\chi$ is stable because a $\mathbb{Z}_2$ symmetry under which $\chi$ is odd remains, due to Lorentz invariance and the fact that it is charged under $U(1)_X$. If instead $m_{A'}<2m_\chi$, a multi-component DM setup arises wherein both the $A'$ and $\chi$ are stable, even though the remnant $\mathbb{Z}_2$, under which $A'$ is odd, is broken. 
Here we 
%will not study the phenomenology of this (very minimal!) multicomponent setup in detail, although it would be certainly worth to do it.
%Instead we 
focus on how DM can be destabilised in the latter framework by extra right-handed neutrinos. 
Adding these seesaw states opens up the possibility of neutrino portal interactions,
\begin{equation}
\delta{\cal L} =  -(Y_L \overline{N_R} \phi \chi_L + Y_{R} \overline{N^c_R} \phi \chi_R +h.c.) \, ,
\label{LagrchiN}
\end{equation}
on top of the usual seesaw interactions,
\begin{eqnarray}
\hspace{-0.2cm}
{\cal L}_\text{seesaw}&=&i \overline{N_R}\partial \hspace{-1.5mm} \slash N_R-\frac{1}{2} m_N (\overline{N_R} N_R^c+\overline{N_R^c} N_R)\nonumber\\
&& -(Y_\nu \overline{N_R} \tilde{H}^\dagger L+h.c.)\, .
\label{seesaw}
\end{eqnarray}
Here we consider only one right-handed neutrino and one SM lepton doublet, $L$. The generalisation to several flavours is straightforward.
%Note that close variants of the model resulting from Eqs.() and () has been considered in \cite{} for various purposes ((((check better what to say here.... talking already here or not about seesaw for $\xi$ etc)))))
In Eq.~\eqref{LagrchiN}, the $Y_{L,R}$ neutrino portal interactions are allowed if the $\phi \chi$ field combination is neutral under $U(1)_X$, so in the following we will assume $Q_\chi=-Q_\phi=1$.\footnote{Any Yukawa interaction, including the SM ones, always requires that the $U(1)$ charges of the particles involved ``miraculously" sum up to 0.}
Note that $Y_L$, $Y_R$, and $Y_\nu$ can all be made real and positive by rephasing appropriately the $\chi_L$, $\chi_R$, and $L$ fields. Note also that the vectorlike character of the new $\chi_{L,R}$ fermions ensures that the model is free of SM and $U(1)_X$ gauge anomalies.

\begin{figure}
\centering
\includegraphics[width=0.65\columnwidth]{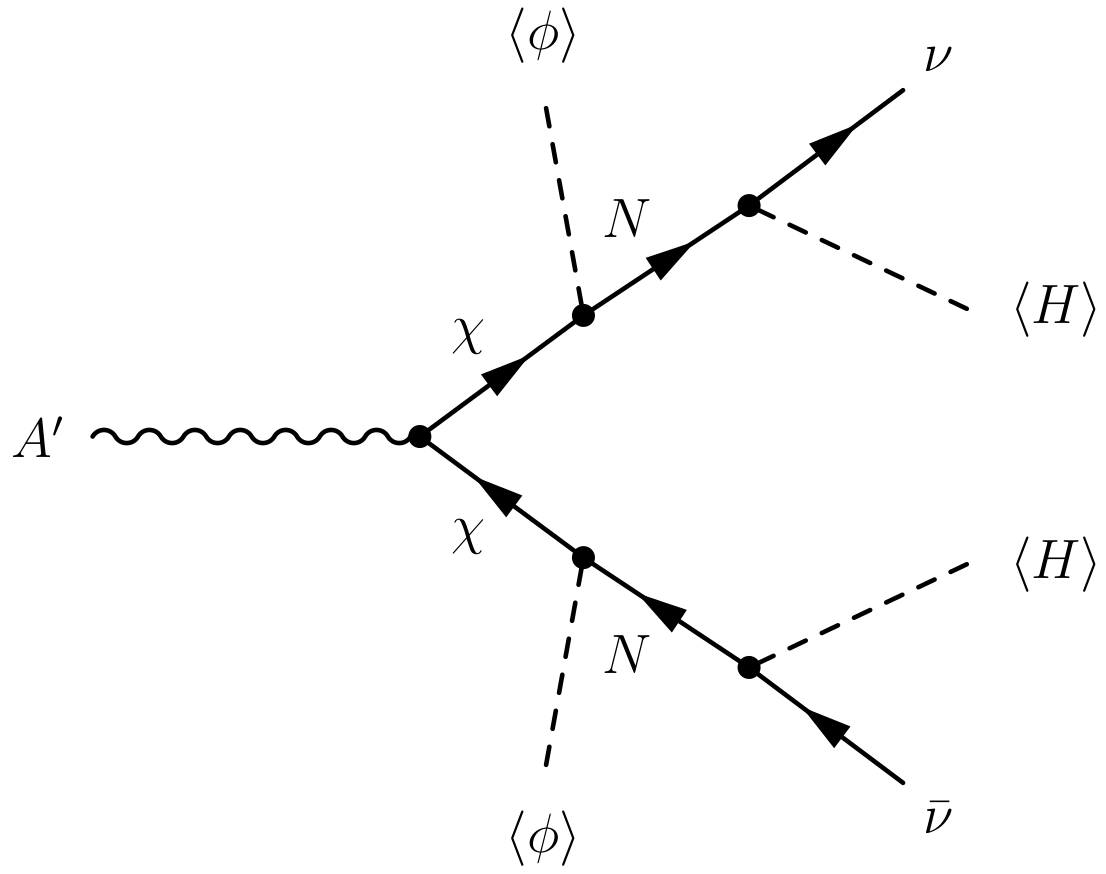}
%\vspace{-0.0cm}
\caption{The $A' \to \nu \bar{\nu}$ decay at tree-level.}
\label{fdAnunu}
\end{figure}

For a heavy right-handed neutrino, where $m_N \gg m_{A'},m_\chi$, the $\chi\rightarrow N \phi$ decays are kinematically forbidden, but $Y_{L,R}$ induces $\chi\rightarrow \nu_L \phi$ decays through seesaw mixing. Similarly, %if $m_{A'}< 2 m_\chi$, 
the $Y_{L,R}$ interactions and seesaw mixing induce $A'\rightarrow \nu_L\bar{\nu}_L$ decays, see Fig.~\ref{fdAnunu}. The amplitude of the first process is suppressed by one power of $m_N$ because it involves one seesaw mixing. The second process instead involves two seesaw mixings and hence is suppressed by two powers of $m_N$.
Thus, the second process can generically lead to neutrino lines with an intensity of order the present experimental sensitivity, whereas the first one gives a lifetime much smaller than the age of the Universe unless $Y_{L.R}$ are tiny.

To compute the decay amplitudes of both processes it is necessary to go to the mass eigenstate basis for the four neutral leptons, $\nu_L, \,N_R,\, \chi_L, \,\chi_R$, and the scalar bosons.  The neutral lepton mass Lagrangian is
\begin{equation}
%\hspace{-0.7cm}
\footnotesize
{\cal L}_\text{mass} = - \frac{1}{2} \begin{pmatrix}
\overline{\nu_{\scaleto{L}{3pt}}^c} & \overline{\chi_{\scaleto{L}{3pt}}^c} & \overline{\chi_{\scaleto{R}{3pt}}} & \overline{N_{\scaleto{R}{3pt}}}
\end{pmatrix} \begin{pmatrix}
0 & 0 & 0 & m \\
0 & 0 & m_\chi & m_{\scaleto{L}{3pt}} \\
0 & m_\chi & 0 & m_{\scaleto{R}{3pt}} \\
m & m_{\scaleto{L}{3pt}} & m_{\scaleto{R}{3pt}} & m_N
\end{pmatrix} \begin{pmatrix}
\nu_{\scaleto{L}{3pt}} \\
\chi_{\scaleto{L}{3pt}} \\
\chi_{\scaleto{R}{3pt}}^c\\ 
N^c_{\scaleto{R}{3pt}}
\end{pmatrix} +h.c.\, ,
\end{equation}
where $m = v Y_\nu/\sqrt{2}$ and $m_{L,R} = v_\phi Y_{L,R}/\sqrt{2}$. 
%For simplicity here we will consider $Y_{L} = Y_R \equiv Y$ so that $\chi_L$ and $\chi_R$ are not split in mass \rupert{[actually there is a small mass splitting of $Y^2 v_\phi^2/m_N + \mathcal{O}(1/m_N^2)$]} and there are only three different mass eigenvalues. In this case, 
The mass eigenstates, $n_i = \begin{pmatrix}
\nu & \chi_1 & \chi_2 & N
\end{pmatrix}^T$, are related to the gauge eigenstates by
\begin{equation}
%\hspace{-1.6cm}
\begin{pmatrix}
\nu_L + \nu_L^c \\
\chi_L + \chi_L^c \\
\chi_R + \chi_R^c \\
N_R + N_R^c
\end{pmatrix} \simeq O\, \begin{pmatrix}
\nu \\
\chi_1 \\
\chi_2 \\
N
\end{pmatrix} \, ,
\end{equation}
with $O$ given by 
\begin{equation}
\begin{pmatrix}
i & - \frac{m(m_L + m_R)}{\sqrt{2} m_\chi m_N} & \frac{im (m_R - m_L)}{\sqrt{2} m_\chi m_N} & \frac{m}{m_N} \\
\frac{i m m_R}{m_\chi m_N} & \frac{1}{\sqrt{2}} + \frac{m_R^2 - m_L^2}{4\sqrt{2} m_\chi m_N} & - \frac{i}{\sqrt{2}} + \frac{i(m_R^2 - m_L^2)}{4\sqrt{2} m_\chi m_N} & \frac{m_L}{m_N} \\
\frac{i m m_L}{m_\chi m_N} & \frac{1}{\sqrt{2}} - \frac{m_R^2 - m_L^2}{4\sqrt{2} m_\chi m_N} & \frac{i}{\sqrt{2}} + \frac{i (m_R^2 - m_L^2)}{4\sqrt{2} m_\chi m_N} & \frac{m_R}{m_N} \\
\frac{-i m}{m_N} & - \frac{m_L + m_R}{\sqrt{2} m_N} & - \frac{i(m_R - m_L)}{\sqrt{2} m_N} & 1
\end{pmatrix} \, ,
\end{equation}
at $\mathcal{O}(1/m_N)$. 
The mass eigenvalues are $\simeq m^2/m_N, m_\chi \mp (m_L \pm m_R)^2/(2m_N)$, and $m_N$, respectively.

For the scalar bosons, after SSB the real scalar of the SM Higgs doublet, $h'$, and hidden sector scalar boson, $\eta'$, mix through the Higgs portal interaction, leading to mass eigenstates,
\begin{equation}
\begin{pmatrix}
h \\
\eta
\end{pmatrix} = \begin{pmatrix}
\cos \varphi & -\sin \varphi \\
\sin \varphi & \cos \varphi
\end{pmatrix} \begin{pmatrix}
h' \\
\eta'
\end{pmatrix} \, ,
\end{equation}
where the mixing angle is
\begin{equation}
\tan 2\varphi = \frac{\lambda_m v v_\phi}{\lambda_\phi v_\phi^2 -  \lambda v^2} \, .
\end{equation}
The mass eigenvalues are
\begin{align}
m_{h,\eta}^2 = \lambda v^2 + \lambda_\phi v_\phi^2 \pm \sqrt{ (\lambda v^2 - \lambda_\phi v_\phi^2)^2 + \lambda_m^2 v^2 v_\phi^2} \, ,
\end{align}
which in the limit of $\lambda_m \ll 1$ reduce to $m_h^2 \simeq 2\lambda v^2$ and $m_\eta^2 \simeq 2 \lambda_\phi v_\phi^2$.

The only two-body final state into which the  hidden vector DM particle can decay at tree-level is a pair of neutrinos.
We find
\begin{equation}
\Gamma(A'\rightarrow \nu\bar{\nu})_\text{tree} \simeq \frac{g_X^2 Y_\nu^4 (Y_L^2 - Y_R^2)^2 v^4 v_\phi^4 m_{A'}}{96 \pi m_\chi^4 m_N^4} \, .
\label{Gammanunu}
\end{equation}
%in the limit that the loop-level contribution is negligible (see below). 
This process is, as we anticipated, suppressed by four powers of the seesaw scale, more precisely by four powers of $(Y_{L,R} v_\phi/m_\chi)(Y_\nu v/m_N)$, as it requires two $\chi\rightarrow N_R \rightarrow \nu_L$ transitions, see Fig.~\ref{fdAnunu}. 
Note that when $Y_L=Y_R$, %the $\chi_{1,2}$ states are mass degenerate and form a Dirac spinor, in case 
the decay width of Eq.~(\ref{Gammanunu}) vanishes at $\mathcal{O}(1/m_N^4)$ as the diagrams with intermediate $\chi_L$ and $\chi_R$ involve a relative negative sign.

When $m_{A'}$ is above the EW scale, many three-body and four-body decays open up by replacing Higgs vev insertions with physical particles in the final state. 
The possible three-body decays are $A' \to \nu \bar{\nu}h$, $A' \to \nu \bar{\nu} Z$, and $A' \to \nu \ell^\pm W^\mp$. 
The allowed four-body decays can easily be deduced. 
Neglecting the final state masses, the rates are
\begin{align}
\Gamma_{A',\text{three-body}} &\simeq \frac{3 g_X^2 Y_\nu^4 (Y_L^2 - Y_R^2)^2 v^2 v_\phi^4 m_{A'}^3}{64 (4\pi)^3 m_\chi^4 m_N^4} \, \label{3body} \\
\Gamma_{A',\text{four-body}} &\simeq \frac{g_X^2 Y_\nu^4 (Y_L^2 - Y_R^2)^2 v_\phi^4 m_{A'}^5}{320 (4\pi)^5 m_\chi^4 m_N^4} \, ,\label{4body}
\end{align}
We see that the phase space suppression compared to the two-body decay is compensated by additional powers of $m_{A'} /v$, so that the three-body rate is larger than the two-body rate for $m_{A'} \gtrsim 2.9$ TeV and the four-body rate becomes dominant for $m_{A'} \gtrsim 12$ TeV. 
On the other hand, replacing $\phi$ vev insertions gives factors of $m_{A'}/v_\phi \lesssim 1$ while paying the price of the phase space suppression, so these decays are subdominant and can be neglected.

\begin{figure}[t]
\centering
\includegraphics[width=0.85\columnwidth]{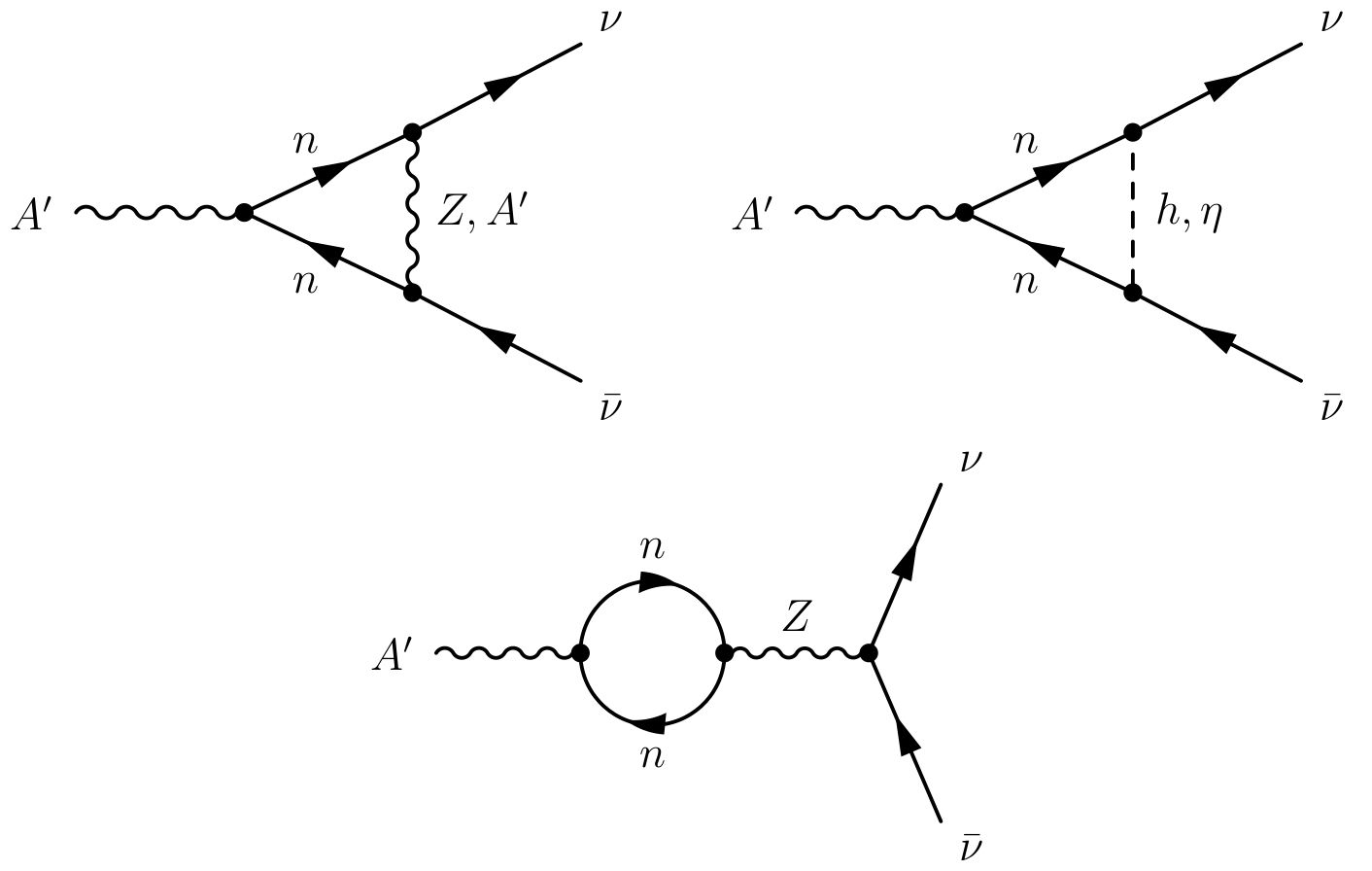}
\vspace{-0.3cm}
\caption{The one-loop diagrams giving $A' \to \nu \bar{\nu}$ decay. 
Here $n = \nu, \chi_{1,2}, N$ are mass eigenstates. }
\label{diagApnn}
\end{figure}

One-loop decay processes also have to be considered. 
The decay to neutrinos, Fig. \ref{diagApnn}, proceeds through the exchange of a scalar or vector boson in the $t$-channel or through one-loop $A'-Z$ mixing. 
The $Z$ exchange diagram dominates, and for $m_{A',\chi}  \gg m_h$, we have
\begin{eqnarray}
\hspace{-0.5cm}
{\cal M}  \simeq & \frac{g_X Y_\nu^2 (Y_L^2 - Y_R^2) v_\phi^2}{128 \pi^2 m_N^2} \log \frac{m_N^2}{m_\chi^2} \overline{u}(p_\nu) \gamma_\mu \gamma_5 v(p_{\bar{\nu}}) \epsilon^\mu(p_{A'}) \, , 
\label{apnnAmplitude}
\end{eqnarray}
plus terms not enhanced by the large log. 
This leads to
\begin{align}
\Gamma & (A'\rightarrow \nu \overline{\nu})_\text{loop} \simeq \frac{g_X^2 Y_\nu^4 (Y_L^2 - Y_R^2)^2 v_\phi^4 m_{A'}}{96 (4\pi)^5 m_N^4} \log^2 \frac{m_N^2}{m_{\chi}^2} \, ,
\label{apnnloop}
\end{align}
when the tree-level contribution can be neglected.\footnote{We will not give here the explicit form of the (constructive) tree-level and one-loop interference term but take it into account in our results below.}
%Alternatively, in the limit $m_\chi \gg m_{A'}$, Eqs. \eqref{apnnAmplitude} and \eqref{apnnloop} are modified by $\log (m_N^2/m_{A'}^2) \to \log (m_N^2/m_\chi^2)$. 
This rate is suppressed by four powers of the $\chi-\nu_L$ mixing and there are no extra powers of $m_N$ in the numerator coming from the fermionic trace or loop integral. Thus, it is of the same order in $1/m_N$ as the contributions of Eqs.~(\ref{Gammanunu})-(\ref{4body}).
Similarly to the point emphasised in \cite{Patel:2019zky}, since the two-body decay is proportional to powers of vacuum expectations values, then for DM masses well beyond the values of these vevs, the one-loop contribution can be greater than the tree-level one. This stems from the fact that the loop contribution involves the propagators of the scalar fields, rather than their vevs. As a result, with respect to the tree-level contributions of Eq.~(\ref{Gammanunu}), the loop factor is compensated by a factor of $m_\chi^4/v^4$. 
For instance, for $m_{A'} = m_\chi$, the rate in Eq. \eqref{apnnloop} is larger than the tree-level width given in Eq. \eqref{Gammanunu} for $m_{A'} \gtrsim 1.6$ TeV. 
Comparing with the four-body decays, we have $\Gamma(A' \to \nu \bar{\nu})_\text{loop}/\Gamma_{A',\text{four-body}} \simeq (10/3) (m_\chi/m_{A'})^4 \log^2(m_N^2/m_{A'}^2)$, hence the two-body decay dominates. Thus the four-body contribution can always be neglected, as can the three-body one.

\begin{figure}
\centering
\includegraphics[width=0.85\columnwidth]{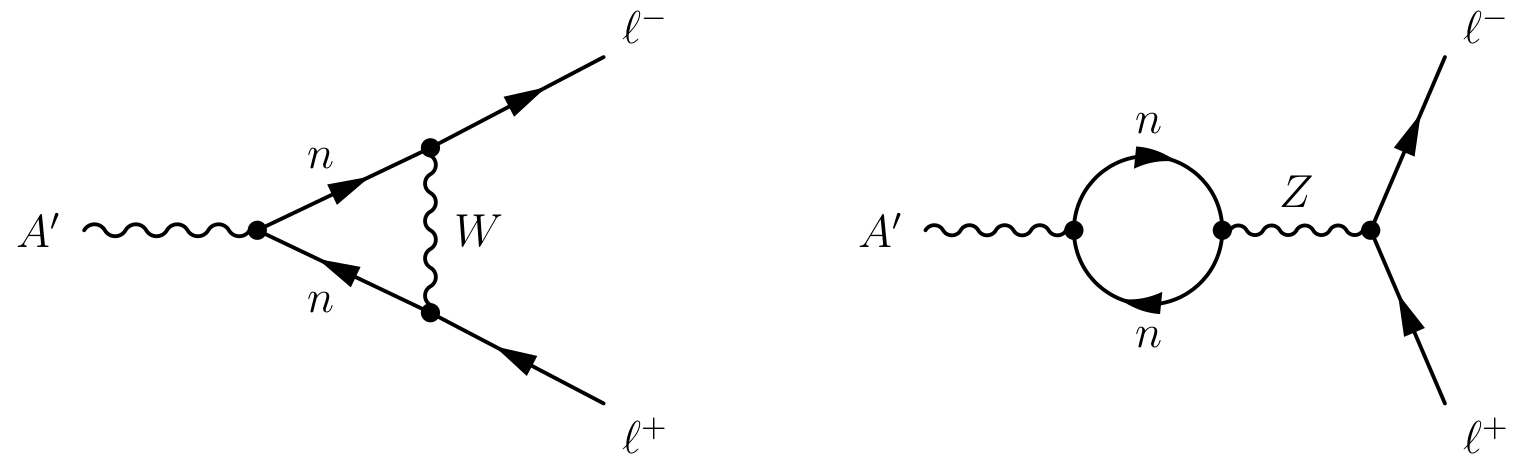}
\vspace{-0.3cm}
\caption{The one-loop diagrams giving $A' \to \ell^- \ell^+$ decay. 
Here $n = \nu, \chi_{1,2}, N$ are mass eigenstates. }
\label{diagApll}
\end{figure}

The decay to charged leptons, shown in Fig. \ref{diagApll}, proceeds either from the exchange of a $W$ in the $t$-channel or through $A'-Z$ one-loop mixing. 
Due to $SU(2)_L$ symmetry, the leading order amplitude for this process is the same as the amplitude for the loop-level decay to neutrinos, (neglecting the final state lepton masses), and hence the partial width $\Gamma(A' \to \ell^+ \ell^-)$ is the same as the width in Eq. \eqref{apnnloop}.

Finally, we note that one-loop decays to bosonic final states, such as $A' \to Z h$ and $A' \to W^+ W^-$, also exist for sufficiently heavy $A'$, with comparable rates to $A' \to \ell^+ \ell^-$. %$A' \to \nu \bar{\nu}$
We will not consider their contributions as they do not bring any spectral features and do not change by much the constraints one can obtain from diffuse fluxes of cosmic rays.

Unlike for the hidden vector decay, the decay of the fermion $\chi$ is suppressed by only two powers of $m_N$ since it involves only one $\chi\rightarrow N\rightarrow \nu$ transition. 
There are many possible decay channels. 
In the limit of $\varphi \simeq 0$, the decay widths to $\eta \nu$ and $h \nu$ are
\begin{equation}
\Gamma(\chi_{1,2} \rightarrow \eta \nu) \simeq \frac{ Y_\nu^2(Y_L \mp Y_R)^2 v^2 m_\chi}{64\pi m_N^2} \left(1 - \frac{m_\eta^2}{m_\chi^2} \right)^2 
\end{equation}
and
\begin{equation}
\Gamma(\chi_{1,2} \rightarrow h \nu) \simeq \frac{Y_\nu^2 (Y_L \pm Y_R)^2 v_\phi^2 m_\chi}{64 \pi m_N^2} \left(1 - \frac{m_h^2}{m_\chi^2} \right)^2 \, .
\label{GammaChiNuH}
\end{equation}

There are also decays to SM gauge bosons, when kinematically allowed, with partial widths
\begin{align}
%\hspace{-0.7cm}
\Gamma(\chi_{1,2} \to W^\pm \ell^\mp) &\simeq \frac{Y_\nu^2 (Y_L \pm Y_R)^2 v_\phi^2 m_\chi}{64\pi m_N^2} f(m_W^2/m_\chi^2) ~, \\
\Gamma(\chi_{1,2} \to Z \nu) &\simeq \frac{Y_\nu^2 (Y_L \pm Y_R)^2 v_\phi^2 m_\chi}{64\pi m_N^2} f(m_Z^2/m_\chi^2) \, ,
\end{align}
where $f(x) = (1-x)^2(1+2x)$. 
Finally, the $\chi$ also decays to $A' \nu$, with partial width
\begin{equation}
\hspace{-0.4cm}
\Gamma(\chi_{1,2} \to A' \nu) \simeq \frac{Y_\nu^2 (Y_L \mp Y_R)^2 v^2 m_\chi}{64 \pi m_N^2} f(m_{A'}^2/m_\chi^2) \, . 
\end{equation}
If $m_\chi<m_\eta, m_{A'}, m_W$ the leading decays are to three SM fermions, mediated by the $W$ or $Z$ boson, which are also suppressed by two powers of $m_N$.

\section{Results}

\begin{figure}[t]
\centering
\includegraphics[width=1.0\columnwidth]{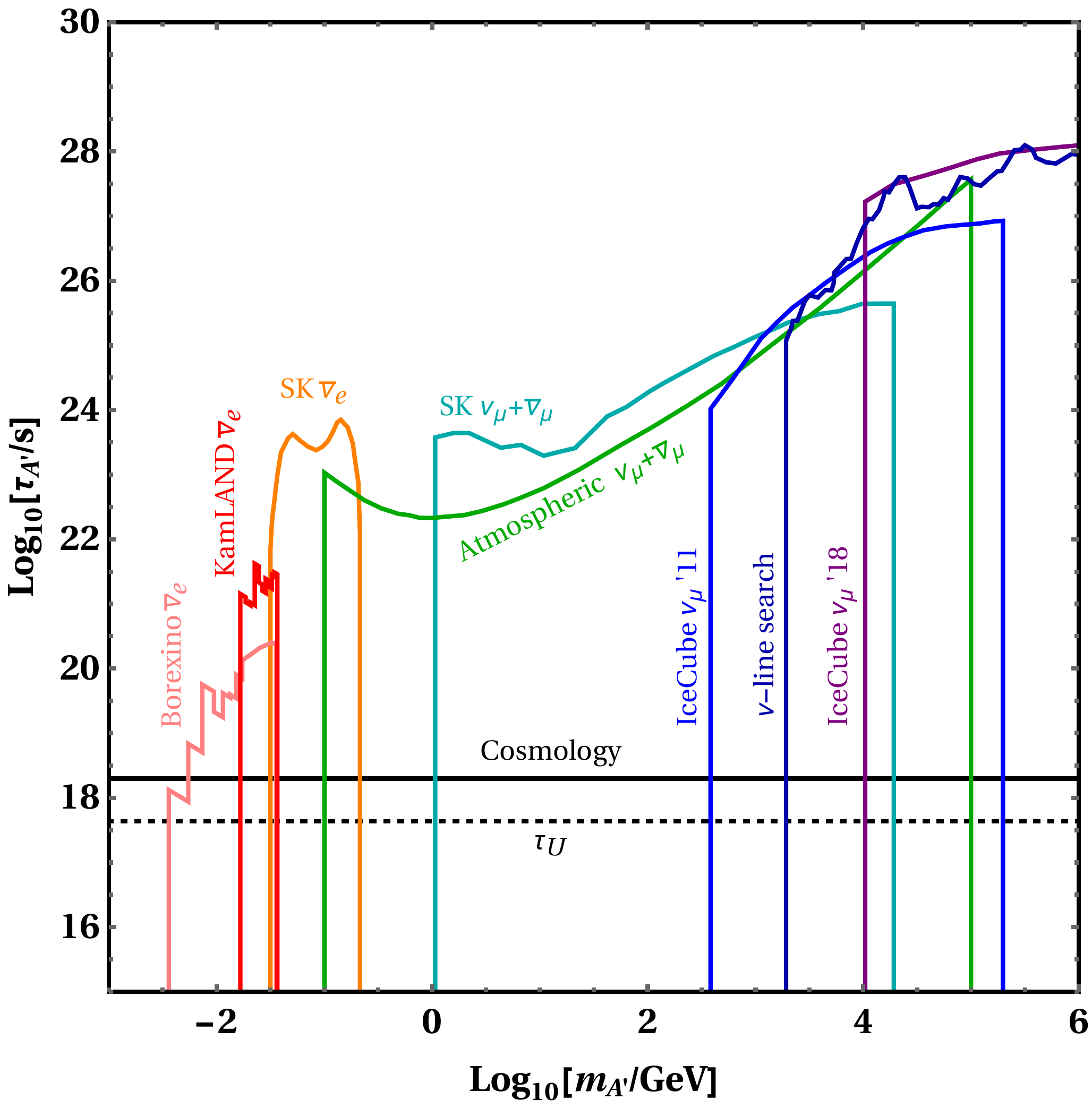}
%\vspace{-0.3cm}
\caption{Bounds on the lifetime of dark matter assuming it decays only into $\nu \bar{\nu}$. Here we assume that the DM couples universally to the three neutrinos flavours.}
\label{Lifetime}
\end{figure}

Experimental constraints on the lifetime of DM decaying into a pair of neutrinos can be found in Refs.~\cite{PalomaresRuiz:2007eu,PalomaresRuiz:2007ry,Aisati:2015vma,FrankiewiczonbehalfoftheSuper-KamiokandeCollaboration:2016pkv,Garcia-Cely:2017oco,Aartsen:2018mxl,Medici:2020wri}. 
The best bounds from DM indirect detection observations are from direct searches of a flux of neutrinos, including those by Borexino \cite{Bellini:2010gn}, KamLAND \cite{Collaboration:2011jza}, Icecube \cite{Aisati:2015vma,Aartsen:2018mxl,Medici:2020wri} and Super-Kamiokande \cite{Malek:2002ns,PalomaresRuiz:2007eu,Zhang:2013tua,FrankiewiczonbehalfoftheSuper-KamiokandeCollaboration:2016pkv}. 
%or from the search of a diffuse flux of cosmic rays emitted by the neutrinos (.....).
Cosmological constraints also exist. Besides the condition that $\tau_{DM}>\tau_U$, CMB data gives $\tau_{DM} > 4.6 \,\tau_U$ \cite{Poulin:2016nat}. 
In Ref.~\cite{Garcia-Cely:2017oco}, many of these constraints were compiled and translated into an upper bound on the $U(1)_{B-L}$ breaking scale as a function of the Majoron mass. Translating them back into constraints on the DM lifetime, and adding the dedicated search for neutrino lines from Icecube data \cite{Aisati:2015vma}, as well as recent IceCube collaboration limits \cite{Aartsen:2018mxl,Medici:2020wri}, Fig.~\ref{Lifetime} shows the various constraints on the DM lifetime. 
The result shown assumes flavour universality, i.e.~$\Gamma(A' \to \nu_\alpha \bar{\nu}_\alpha)$ is the same for $\alpha = e,\mu,\tau$ (in this case the neutrino mass hierarchy plays no role). Modifying the branching ratios to each flavour or the neutrino mass hierarchy only mildly affects the results.

\begin{figure}
\centering
\vspace{0.14cm}
\includegraphics[width=1.0\columnwidth]{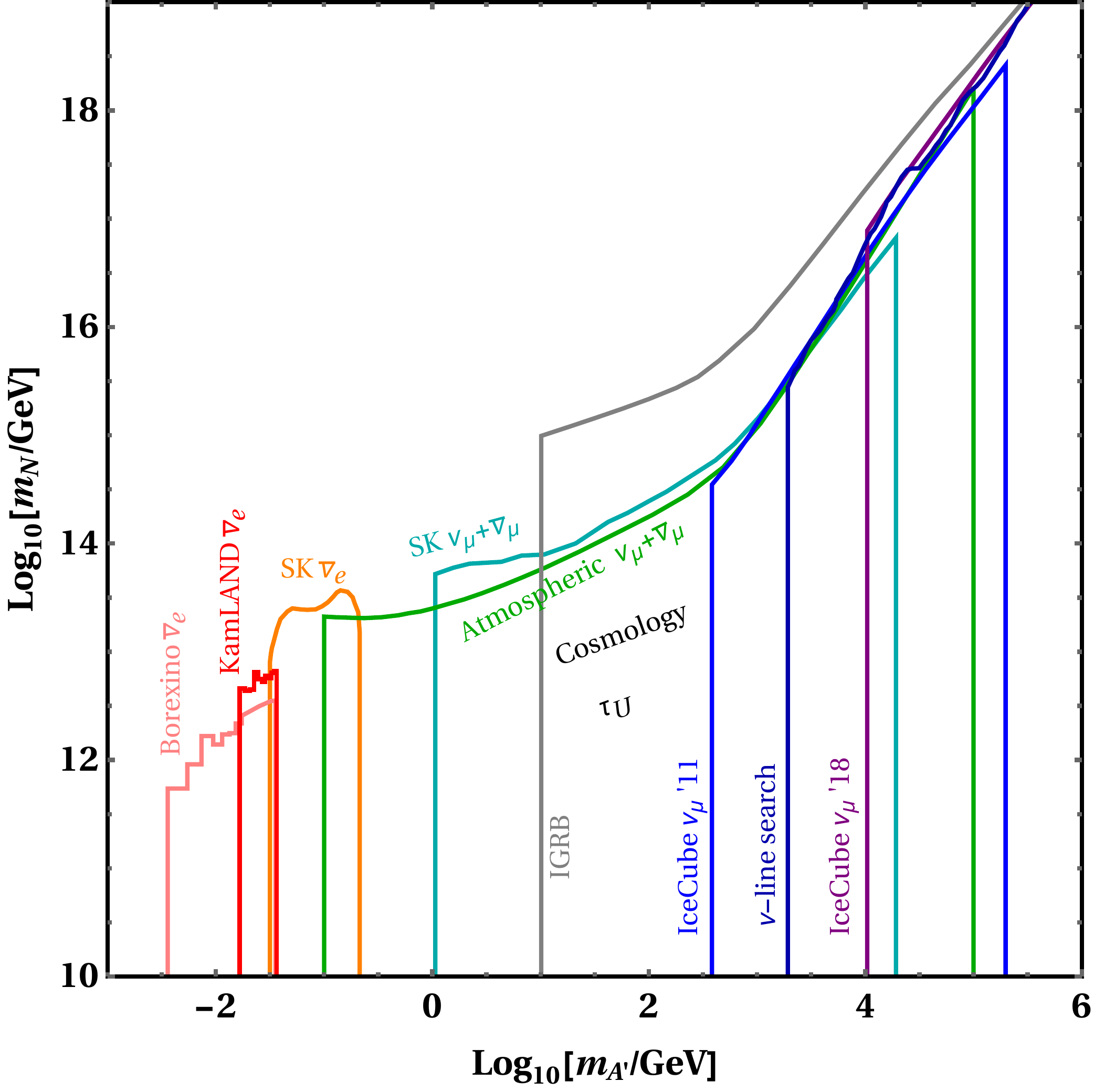}
\caption{Lower bounds on the heavy sterile neutrino mass, $m_N$, from constraints on the lifetime of DM which decays as $A' \to \nu \bar{\nu}$ and $A' \to \ell^+ \ell^-$. Here we assume that the couplings are of order unity and $m_\chi=m_{A'}$, and that the DM couples universally to the three flavours.}
\label{mNbound}
\end{figure}

In the following, to present the results, we will take a simple benchmark case where the couplings are equal to %of order
 unity, $g_X = Y_\nu = |Y_L^2 - Y_R^2| = 1$, and $m_\chi = m_{A'}$. 
This implies $m_{DM}=v_\phi$ and gives $\Gamma(A'\rightarrow \nu\bar{\nu})= (1/96\pi)\cdot v^4 m_{A'}/m_N^4$ at tree-level.
Fig.~\ref{mNbound} gives, for this straightforward case, the lower bound on $m_N$ we get from the various constraints on the lifetime in Fig.~\ref{Lifetime}. 
Again, it is assumed that the DM decays in a flavour-universal way. 
As expected, the values are typically of order the Weinberg operator scale when $m_{DM}$ is of order the electroweak scale. 
Of course, nothing guarentees that $m_\chi$ must necessarily be of order $v_\phi$, since these two scales are independent in the setup we consider.
%Thus, one can play with the $v_\phi/m_\chi$ ratio. 
For $v_\phi\neq m_\chi$ and small $m_{A'}$, such that the tree-level part of the $A' \to \nu \bar{\nu}$ amplitude dominates, the bounds on $m_N$ have to be simply rescaled by one power of the $v_\phi/m_\chi$ ratio (assuming still $m_{A'} < 2 m_\chi$). 
For larger $m_{A'}$, when the one-loop contribution dominates, the bound on $m_N$ depends only logarithmically on $v_\phi/m_\chi$. 
%If one assumes that the $\chi$ state is not larger that a few TeV or say few tens of TeV (as the DM particle) the limits of Fig.~\ref{mNbound} typically holds for couplings of order unity. 
Also, considering couplings smaller than unity clearly leads to a less stringent lower bound on $m_N$ than in Fig.~\ref{mNbound}.

Fig.~\ref{mNbound} also shows the lower bound we get on $m_N$ from $A' \to \ell^+ \ell^-$ (with $l=e,\mu, \tau$), using the results of \cite{Blanco:2018esa} obtained from Fermi-LAT data  \cite{Ackermann:2014usa} on the isotropic gamma-ray background (IGRB). %\footnote{Limits on $\tau_{DM}$ from decays to bosonic final states, i.e. decays to $W^+ W^-, ZZ, Zh, hh$, are also given in \cite{Blanco:2018esa}, and are almost identical to those from decays to charged leptons.}
Comparing this bound with the ones from the neutrino channel, one observes that at the moment charged lepton limits are more stringent for $m_{DM}\gtrsim 10$~GeV
by a factor of a few (although this relative factor depends somewhat on the flavour composition of the DM decays and the neutrino mass hierarchy). This is interesting because it means that improving the limits for the neutrino channel by a factor of a few would open the possibility of seeing both an associated flux of neutrinos and charged leptons. As mentioned above, for $m_{DM}\gtrsim 1.6$ TeV the loop contribution dominates and predicts an equal decay width for both channels (similarly to the setup of \cite{Patel:2019zky}).

\begin{figure}%[t]
\centering
\includegraphics[width=1.0\columnwidth]{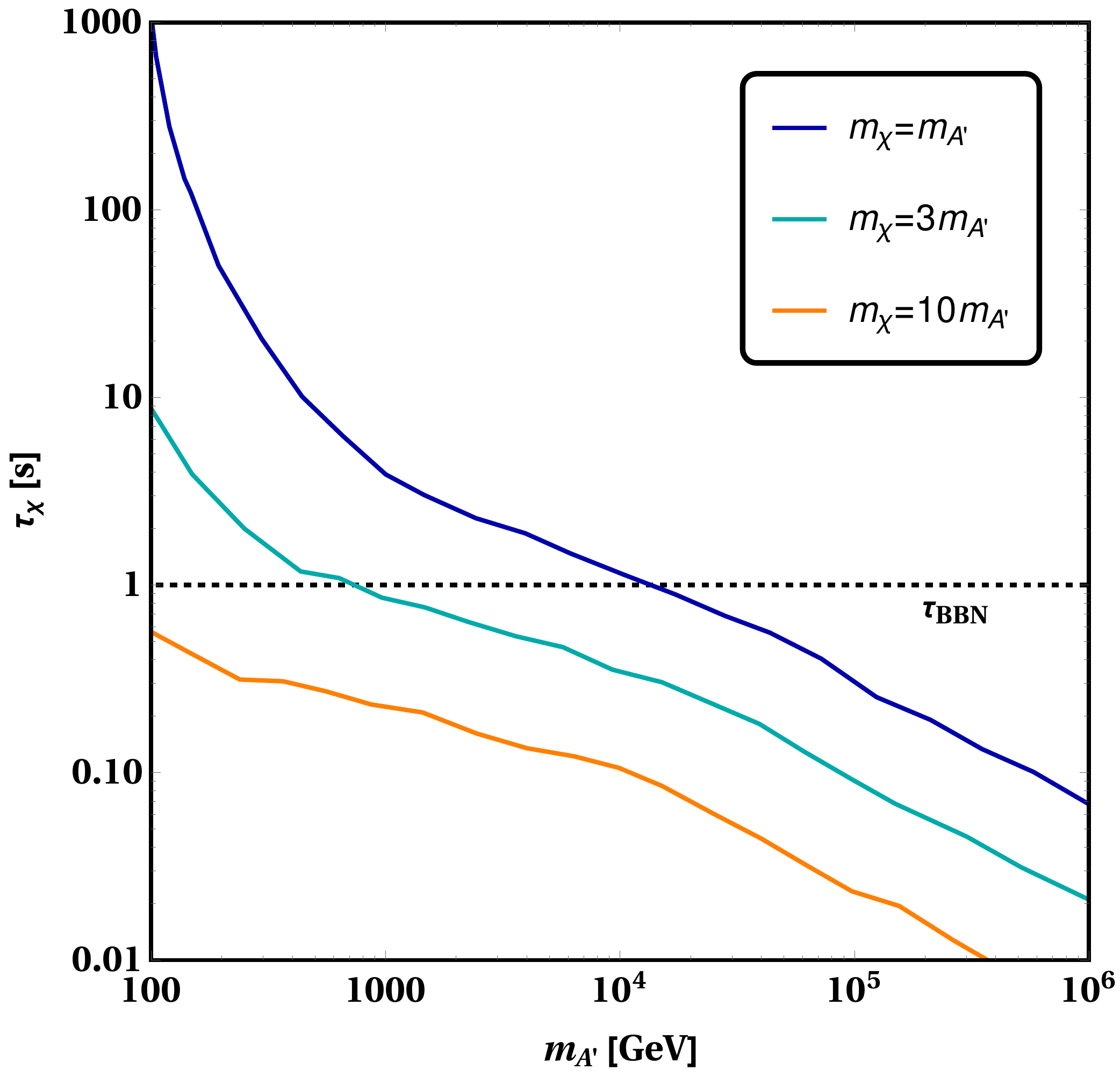}
%\vspace{-0.3cm}
\caption{Lifetime of the $\chi_{1,2}$ for different ratios of $m_\chi/m_{A'}$, given couplings equal to unity and the lowest value of $m_N$ allowed by experiments.}
\label{chiLifetime}
\end{figure}

If the doubly seesaw-suppressed decay width of the $A'$ is of order the experimental sensitivity,
the $\chi$ lifetime is expected to be much smaller than the age of the Universe, 
since the corresponding decay width of $\chi$ is only singly suppressed by the seesaw scale. 
%(i.e~suppressed by only two powers of the seesaw mixing), 
In Fig.~\ref{chiLifetime} the dark blue line gives the lifetime of $\chi_{1,2}$ we get assuming the same benchmark set of parameters as
for Fig.~\ref{mNbound}.
%couplings equal to unity (specifically, $g_X = Y_\nu = |Y_L \pm Y_R| = 1$) and $m_N$ as given in Fig.~\ref{mNbound}. 
We restrict ourselves to $m_\chi > 100$ GeV, so that the $\chi$ has kinematically allowed two-body decays.
%, and consider different values of $m_\chi/m_{A'}$. 
As the figure shows, its lifetime is around the age of the Universe at the beginning of the BBN epoch, $\tau_{\chi} \sim 1$~sec.  Since the $\chi$ decay produces electromagnetically coupled SM particles, BBN typically requires that the lifetime must be smaller than 1~sec. 
%This constraint is largely relaxed if the relative abundance of $\chi$ is much smaller than the one of the $A'$ (see e.g.~\cite{Hufnagel:2018bjp}).
%To have a suppressed $\chi$ abundance is in fact easy to achieve here, as soon as $m_\chi$ is larger than $m_{A'}$, since in this case  the $\chi$ abundance will be Boltzmann suppressed from annihilating into \textbf{pairs} %a pair of $A'$.

It is interesting to note that the ratio of the lifetime of DM allowed by indirect detection, $\tau_{DM}\gtrsim 10^{26-29}$~sec, and the age of the Universe at BBN time, $t_{BBN}\sim 1$~sec, is rather similar to  the ratio of the neutrino mass scale, $m_\nu\sim 0.1$~eV, and the seesaw scale, $\Lambda\sim 10^{15}$~GeV, which is $m_\nu/\Lambda\sim 10^{-25}$. 
This means that if the decay width of the hidden vector is of order the experimental sensitivity for neutrino lines, being suppressed by four powers of the seesaw scale, particles whose decay is suppressed by two powers of the seesaw scale can have already disappeared by the time of BBN. 
The ratio of the lifetimes (when the tree-level $A'$ decay dominates) scales as $\tau_\chi/\tau_{A'} \sim (Y_\nu v/m_N)^2 (g_X Y_{L,R} v v_\phi/m_\chi^2)^2 (m_{A'}/m_\chi) \sim (m_\nu/m_N)\cdot C$ with $C= (g_X Y_{L,R} v v_\phi/m_\chi^2)^2 (m_{A'}/m_\chi)$. 
%The last factor can easily bring an extra suppression (if necessary), as demonstrated in Fig.~\ref{chiLifetime}.
To illustrate the above, one can also write down the lifetime of the $\chi$ from $\chi_{1,2}\rightarrow SM$ channels, in the $m_\chi \gg v$ limit, as\footnote{Another option is to assume tiny values of $Y_L$ and $Y_R$, so that the lifetime of $\chi$ is larger than the age of the Universe and also larger than indirect detection lower bounds on the lifetime. In this case the direct connection between the seesaw scale and DM lifetime is lost (and moreover this renders the lifetime of $A'$ unobservably long).}
\begin{equation}
%\hspace{-0.5cm}
\tau_{\chi_{1,2}} \simeq \frac{1\text{ sec}}{Y_\nu^2 (Y_L \pm Y_R)^2} \left( \frac{1.5 \text{ TeV}}{m_\chi^{1/3} v_\phi^{2/3}} \right)^3% \left(\frac{1.5 \text{ TeV}}{v_\phi} \right)^2
 \left(\frac{m_N}{10^{16} \text{ GeV}} \right)^2 \, .
\end{equation} 
As Fig.~\ref{chiLifetime} shows, if $m_\chi=m_{A'}$, the lifetime of $\chi$ is smaller than one second only if $m_{A'}\gtrsim 20$~TeV. However the C factor above can easily be reduced by decreasing couplings and/or increasing $m_\chi$ with respect to $m_{A'}$. As an example, Fig.~\ref{chiLifetime} also gives the lifetimes keeping couplings equal to unity but taking $m_\chi > m_{A'}$ and still taking the lowest value of $m_N$ allowed by indirect detection experiments.
It shows that in this case values of $m_{A'}$ of order the electroweak scale or below quickly become compatible both with observable neutrino and $\gamma$ fluxes and with BBN.\footnote{Actually, the $\tau_\chi\lesssim 1$~sec BBN constraint ought to be applied if the $\chi$ particles are numerous, as would be the case if, for instance, they decouple from the thermal bath relativistically. However,
we would instead expect the $\chi$ abundance to be of the same order as the $A'$ one, as annihilations of both particles can be dominantly driven (and Boltzmann suppressed) by the same $U(1)_X$ gauge interactions. In this case, BBN allows for $\chi$ lifetimes a few orders of magnitude larger than 1 sec \cite{Hufnagel:2018bjp}, 
and the BBN constraint is very easily satisfied, see Fig.~\ref{chiLifetime}.}
%so the BBN constraints are satisfied more easily than our requirement of a lifetime shorter than 1 sec.
%saturating the experimental sensitivity on the observation of 

%\rupert{[I suppose we drop the following paragraph]\\}
%In Fig.~CCC we present the neutrino energy spectrum we get summing the 2-body $\nu \bar{\nu}$ channel and the 3 and 4 body ones. It is shown as must be observed by a detector with say 20\% energy resolution.
%For $m_{DM}$ below ??? the 2 body channel dominate the total decay width and the line is still clearly observable. This remains true up to ? TeV.
%Above this value, the branching ratio is so much dominated by the 3 body channels that the line becomes to be difficult to observe. One concludes therefore than a spin-1 vector DM setup can lead to a neutrino line with an intensity naturally of order the expected sensitivity provided $m_{DM}<???$.{\bf ((((Do we plot that or not. I guess not))))}

\section{Comparison with other seesaw induced DM decay setups}

Concerning the decays, the main difference between the setup we consider and other scenarios where a DM decay is also induced through the seesaw interactions is that the decay width into a pair of charged lepton is suppressed by four powers of the seesaw scale rather than by two.
As is well known, the Majoron, i.e.~the pseudoscalar DM candidate coupling to $N\overline{N^c}$, not only decays into a pair of neutrinos with a width suppressed by four powers of the heavy $m_N$ scale, but also into a pair of charged leptons, $\ell^+ \ell^-$, at the one-loop level with a width suppressed by only two powers of $m_N$ (from s-channel Z exchange and t-channel $W$-exchange diagrams, similar to those in Fig.~\ref{diagApll}) \cite{Pilaftsis:1993af,Frigerio:2011in,Garcia-Cely:2017oco}.
This is the result of the chirality flip required in the Majoron case, bringing an extra $m_f^2 m_N^2/m^4_{W,Z}$ factor in the decay width. 
% so that the decay amplitude into a charged lepton pair is suppressed alltogether by only one power of $m_N$ rather than 2. 
Although the width is greatly suppressed by the loop factor and the square of the small charged lepton mass, it still leads to much too fast a decay unless one takes the Majoron to be rather light (below $\sim 100$~MeV) and/or we assume that the Yukawa coupling, $Y_N$, which leads to the masses of the right-handed neutrinos, is tiny, which implies $m_N\ll 10^{15}$~GeV%or, equivalently, $Y_\nu \ll 1$
.\footnote{The interaction coupling the Majoron to a pair of right-handed neutrinos also leads to the right-handed neutrino masses, ${\cal L}\owns -Y_N \phi \overline{N^c } N \owns -Y_N f {N^c N}$ with $\phi= (f+\eta) e^{i\theta/f}$, $\langle |\phi|\rangle=f$, and $\theta$ the Majoron.
As a result,  the decay width of the Majoron into a pair of charged leptons, which typically scales as $(m_\nu/v)^2(m_N/f)^2 (m_f/v)^2 m_\theta$, doesn't decrease when $m_N$ increases for fixed neutrino masses.
%, so the lighter the $N$, the slower the decay. 
Conversely, in our setup the width of $A'$ decreases when $m_N$ increases, for fixed neutrino masses.}
%, so that the lighter the $N$, the faster the decay.}
Here, instead, all decays are suppressed by four powers of the large scale $m_N$ and are therefore naturally enough suppressed, even for much larger DM masses.\footnote{Similarly, this is different from the decay of a $Z$ into a $b \bar{b}$ pair with a heavy top quark pair and a $W$ in the loop \cite{Bernabeu:1987me}, which displays two powers of the top quark masses in the amplitude, due to the fact that for large momentum in the loop, the longitudinal $W$ exchange implies two powers of the top Yukawa couplings. In the model we consider, the $W$-exchange diagram of Fig.~\ref{diagApll} in the large momentum limit instead implies two powers of $m_{A'}$. This explains why the loop decay width is enhanced by $m_{\chi}^4/v^4$ relative to the tree-level width.} Thus, the production of observable energetic neutrino lines is achieved in a more straigthforward manner than for the Majoron case. For an analysis of neutrino line searches from Majoron decay, see \cite{Garcia-Cely:2017oco}.

The possibility of having a slow, seesaw-induced decay of a vector gauge boson was also studied recently in Ref.~\cite{Patel:2019zky}. 
The model also considers an extra  $U(1)'$ gauge structure, spontaneously broken by the vev of an extra scalar, with the $U(1)'$ gauge boson being the DM particle. 
It also involves a neutrino portal interaction involving a singlet right-handed neutrino and an extra scalar and fermion, both charged under the $U(1)'$. As mentioned above, in \cite{Patel:2019zky} it was emphasised that the loop contributions to the DM decay width dominate for large DM masses. This setup therefore also leads to characteristic neutrino lines from DM decays into a pair of neutrinos.
The model nevertheless differs from the one we studied above in various ways.
Firstly, it assumes two sets of right-handed neutrinos rather than one, a ``visible'' set coupling to the SM doublet of leptons (in the usual seesaw Yukawa way) and a ``hidden sector'' set coupling to the extra charged fermion and charged scalar. 
The two sets mix through tiny off-diagonal Majorana mass terms. 
Secondly, the extra charged fermion is chiral, rather than vector-like,  and acquires its mass through a seesaw mechanism in the hidden sector.
The seesaw-induced extra fermion mass is assumed not to be tiny, thereby requiring right-handed neutrinos with masses much below the Weinberg operator scale. 
Since the extra fermion obtains its mass via a seesaw mechanism, the one-loop induced widths of DM decays into pairs of neutrinos or charged leptons is suppressed by only two powers of the right-handed neutrino masses (rather than four as above). 
All this leads to too rapid a decay of the DM unless there is some tiny parameter entering into play.
This is achieved by assuming that the mass mixing between the sets of right-handed neutrinos is very small. 
Thirdly, the chiral structure assumed requires the existence of extra fermions charged under the $U(1)'$ in order to cancel gauge anomalies.

\section{Non-abelian case}

Instead of the abelian hidden sector gauge structure above, one could have considered a non-abelian symmetry as well. The simplest possibility is a $SU(2)_X$ gauge structure, as in \cite{Hambye:2008bq}. In this case, this gauge symmetry is broken by a complex scalar doublet and one is left with a degenerate triplet of DM gauge bosons protected by the remnant custodial symmetry. The Lagrangian is the same as for the abelian case, Eq.~(\ref{hiddenvectorLagr}), provided that now $\phi$ is the doublet and the $F_{\mu\nu}$ field strength and covariant derivative stand for the $SU(2)_X$ ones. Such a structure can also couple to the seesaw states provided that the vectorlike fermion, $\chi$, is now a doublet, in which case Eq.~(\ref{LagrchiN}) also holds. The DM decay phenomenology 
is essentially the same as for the abelian case. If $m_\chi>m_{DM}/2$, the non-abelian gauge bosons do not decay to a pair of $\chi$ fermions and are destabilised only by seesaw-suppressed interactions, just as in the abelian case (up to $SU(2)_X$ combinatorial factors of order unity). In the non-abelian case there is no possibility of kinetic mixing, so that one does not need to assume that this mixing doesn't exist in order to avoid the associated fast decay.

\section{Summary}

If in a new sector a fermion singlet combination of dimension 5/2 can be written down, i.e.~a ``$\chi \phi$'' singlet bilinear,
this sector can couple to the SM through a neutrino portal interaction, $ \bar{N}\chi\phi$.
This induces a $\chi-\nu$ mixing mediated by a right-handed neutrino, $N$.
If the DM particle in this new sector couples to the  $\chi$ fermion, it can eventually decay into a final state containing ordinary neutrinos. 
%When this decay is of the two-body type, it 
This can lead to the emission of a striking neutrino line that can be searched for.
The decay width in this case is necessarily suppressed by powers of the seesaw scale, i.e.~by powers of the most experimentally motivated UV physical scale that we know of at the moment. 
This nevertheless involves the non-trivial requirement that the finite DM lifetime induced in this way is not too short, and in particular is of the order of the present experimental sensitivity. To this end, known setups of this kind typically require an additional large, ad hoc (coupling) suppression of the DM width. In this letter, we have presented examples of setups where this can be avoided, so that a DM lifetime of order the experimental sensitivity can be entirely associated with the largeness of the Weinberg operator scale, and nothing else. Given the similarity between the Weinberg operator and GUT scales, this offers the interesting possibility of probing UV physics at scales as high as the GUT scale. 
These results are characteristic of spin-1 DM scenarios, as considered above. Instead, a scalar or fermion DM particle gives in the simplest realisations (such as in the Majoron model or the example of the $\chi$ decay above) a lifetime suppressed by only two powers of the seesaw scale. Besides predicting neutrino lines, the spin-1 setups also predict, for large DM masses, an equal production of pairs of charged leptons.
%It is to be anticipated that the results obtained here would be valid for any other DM setup whose lifetime is suppressed by 4 powers of the seesaw scale. For a scalar DM particle this is nevertheless 

% $\phi_{DM} \rightarrow \phi\phi^\dagger\nu \bar{\nu}$ through one loop diagram with $\chi$ exchange in t-channel....)\\

\section*{Acknowledgments}
This work is supported by the ``Probing dark matter with neutrinos" ULB-ARC convention, by the F.R.S./FNRS under the Excellence of Science (EoS) project No. 30820817 - be.h ``The H boson gateway to physics beyond the Standard Model'', by the FRIA, and by the IISN convention 4.4503.15.

\bibliographystyle{apsrev4-1}
\bibliography{seesawDMdecay.bib}

\end{document}